\documentclass{llncs}

\usepackage{makeidx}
\usepackage{amssymb}
\usepackage{listings}
\usepackage{indentfirst}
\usepackage{verbatim}
\usepackage{amsmath, amssymb}
\usepackage{graphicx}
\usepackage{color}
\usepackage[belowskip=-15pt,aboveskip=0pt]{caption}

\definecolor{light-gray}{gray}{0.90}
\newcommand{\graybox}[1]{\colorbox{light-gray}{\texttt{#1}}}

\lstdefinelanguage{ocaml}{
keywords={let, begin, end, in, match, type, and, fun, function, try, with, class, 
object, method, of, rec, repeat, until, while, not, do, done, as, val, inherit, 
new, module, sig, deriving, datatype, struct, if, then, else, ostap, open, generic, virtual},
sensitive=true,
basicstyle=\small,
commentstyle=\small\itshape\ttfamily,
keywordstyle=\ttfamily\underbar,
identifierstyle=\ttfamily,
basewidth={0.5em,0.5em},
columns=fixed,
fontadjust=true,
literate={->}{{$\to$}}1,
morecomment=[s]{(*}{*)}
}

\begin{document}

\mainmatter

\title{Code Reuse With Transformation Objects}

\author{Dmitri Boulytchev}
\institute{St.Petersburg State University,\\ 
Saint-Petersburg, Russia\\
\email{dboulytchev@math.spbu.ru}}

\maketitle

\begin{abstract}
We present an approach for a lightweight datatype-generic programming in Objective Caml 
programming language\footnote{http://caml.inria.fr} aimed at better code reuse. We show 
that a large class of transformations usually expressed via recursive functions with
pattern matching can be implemented using the single per-type traversal function and 
the set of object-encoded transformations, which we call \emph{transformation objects}. 
Object encoding allows transformations to be modified, inherited and extended in a 
conventional object-oriented manner. However, the data representation is kept untouched 
which preserves the ability to construct and pattern-match it in the usual way. 
Our approach equally works for regular and polymorphic variant 
types~\cite{PolyVar} which makes it possible to combine data types and their transformations
from statically typed and separately compiled components. We also present an
implementation which allows to automatically derive most functionality from a slightly 
augmented type descriptions.
\end{abstract}

\section*{Introduction}

Statically typed functional languages are widely renowned for the gears they provide
for describing complex data structures and their transformations. One of the most 
utilized tools are algebraic data types (ADTs) which allow to construct graph-shaped
data structures and inspect them by matching against patterns. Parametric polymorphism 
makes it possible to apply transformation functions for various concrete versions of 
described data structure thus facilitating massive code reuse. However transformations 
for regular ADTs suffer from the lack of extensibility: there is no easy way to 
modify/update their behavior without the complete reimplementation (at least for the 
recursive ADTs, which are the most important case). In contrast, in the parallel
universe of object-oriented programming the specialization or reuse of behavior for the 
cases of interest is a matter of general practice; however, a certain price in the form 
of overweighted object representation and mixing of code and data is paid on that way. 
The winning combination of both approaches --- polymorphic ADTs \emph{and} their 
extensible transformations --- looks quite promising for the languages which combine 
functional and object-oriented constructs.

We present a framework for Objective Caml\footnote{https://code.google.com/p/generic-transformers} which is based on object representation of transformations. Object encoding allows 
transformations to be inherited, modified and reused; transformation objects contain no 
data, which provides perfect code and data separation and preserves all other means of 
manipulation for the regular ADTs. 

The set of transformations expressible using our approach can easily be characterized in 
terms of classic cycle-free attribute grammars~\cite{AGKnuth}. If we interpret the data 
structure as a derivation tree, the result of the transformation as a synthesized attribute 
and all additional parameters of the transformation as inherited attributes then any 
transformation can be implemented using our framework if no attribute value depends on 
itself.

Another interesting feature of the approach is that it provides a clear and well-established
interface between generic and specific code. Thus the framework we describe is in turn 
becomes extensible. We equip it with a plugin system which implements in a user-defined 
manner the similar functionality as ``deriving'' primitive in Haskell.

Our solution is completely type-driven which allows automatic implementation. 
We implemented our approach in the form of \lstinline{camlp5}\footnote{http://pauillac.inria.fr/\textasciitilde ddr/camlp5} syntax extension and runtime support library.

All code snippets in the following sections are written in Objective Caml. While we 
strived to make all examples as much self-contained as possible a certain familiarity 
with Objective Caml fundamentals is still desirable, especially with those concerning 
object-oriented constructs and types. 

Most of the examples in the following sections refer to the transformations and 
tasks typical for the domain of programming languages implementation and compilers 
since this is the area of our interest. We believe, however, that the proposed 
functionality is generic enough to be used in other areas as well.

\section{Transformation Objects: an Overview}
\label{overview}

In this section we present transformation objects as a programming pattern, provide
some motivations and describe the proposed solution in a nutshell.

Consider the type of lambda expressions and its visualization function --- 
\lstinline{show}:

\begin{lstlisting}
    type lam = 
      Var   of string 
    | App   of lam * lam
    | Lam   of string * lam

    let rec show = function
    | Var   s     -> "Var (" ^ s ^ ")"
    | App  (x, y) -> "App (" ^ show x ^ ", " ^ show y ^ ")"
    | Lam  (x, y) -> "Lam (" ^ x ^ ", " ^ show y ^ ")"
\end{lstlisting}

Sometimes (actually, more often than one would desire) the behavior
of such a function has to be slightly modified --- for example, one
may need to show ``\lstinline{Var s}'' as just ``\lstinline{s}'', omitting
the constructor name. Since the modification is quite modest, we might 
expect a significant code reuse. The na\"ive modification

\begin{lstlisting}
    let better_show = function
    | Var s -> s
    | x     -> show x
\end{lstlisting}

\noindent however, would not work as we desired --- we resorted to the
``old'' \lstinline{show} which was left unmodified. Thus our modification will work
only for the top-level \lstinline{Var}s. The only remaining 
option is to rewrite \lstinline{show} completely which is at least regretful
and at most impossible (for example, when we deal with an external
library function with hidden implementation).

The desired extensibility can easily be provided by the conventional object-oriented
encoding: we can represent expressions using a hierarchy of classes (one class per a
constructor) and provide in each class corresponding method \lstinline{show}. 
However, thus we would lose the ability to match expressions against patterns; moreover, 
object data representation would force us to implement virtually every expression-processing
function as a set of per-class methods scattered through various class definitions. 
Apart of being much more verbose this solution would be more error-prone and
less readable.

The better solution would be to apply object-oriented encoding to the \emph{transformation}
itself, not to the transforming \emph{data}. So \emph{transformation objects} come into
consideration. Transformation object is just an object which contains some functionality
needed to perform the transformation. If we deal with the algebraic data types then the
appropriate representation for a transformation object for a given type is a
collection of per-constructor transformation methods. Besides that we need a function
which performs a pattern-matching on the transforming data structure and dispatch the 
control to the transformation object's methods. Taking these considerations into account 
we may implement our \lstinline{show} function in the following manner:

\begin{lstlisting}
    let rec generic_show t e = 
      let self = generic_show t in
      match e with
      | Var   s     -> t#c_Var s
      | App  (x, y) -> t#c_App self x y
      | Lam  (x, y) -> t#c_Lam self x y

    class show_lam = object
      method c_Var  s                     = "Var (" ^ s ^ ")"
      method c_App (f : lam -> string) x y = "App (" ^ f x ^ ", " ^ f y ^ ")"
      method c_Lam (f : lam -> string) x y = "Lam (" ^ x ^ ", " ^ f y ^ ")"
    end

    let show e = generic_show (new show_lam) e
\end{lstlisting}

Here \lstinline{generic_show} is a top-level function parameterized by the transformation 
object \lstinline{t}. It exhaustively matches the expression against patterns
and dispatches the control to the appropriate transformation object methods. 
Note that some of these methods have to be additionally parameterized by the 
transformation function in question (\lstinline{self}) since we do not want them 
to be directly dependent on \lstinline{generic_show} (this would compromise the very 
idea of extensibility).

Now the customized version of \lstinline{show} can really be implemented in a reusable 
(albeit a bit verbose) manner:

\begin{lstlisting}
    let better_show e = generic_show (object inherit show_lam
                                        method c_Var s = s
                                      end) e
\end{lstlisting}

While this time we managed to provide an appropriate implementation the generality
of the described approach is still unclear. Do we need to implement both the traversal
function and transformation-specific class each time we need an extensible solution? 
Can this process be automated in a type-driven manner? 

The key observation we can make so far is that the traversal function 
(\lstinline{generic_show}) in our example turned out to be more generic that we 
expected --- it contains no specific ``show'' functionality apart from the transformation
object it takes as a parameter. In the following section we describe a generalized
version of presented pattern which allows to systematically produce transformations 
from type descriptions. Each such transformation is expressed using a single per-type 
traversal function which is generated by the framework as well.

\section{Type-Driven Transformation Objects}
\label{elaborated}

In this section we describe a per-type abstract \emph{attribute} transformer, 
type-specific traversal function and some auxiliary notions needed for our approach 
to work. We call our transformers \emph{attribute} since their design was inspired by 
the notion of computations defined by attribute grammars~\cite{AGKnuth}. However, 
we do not use attribute grammar formalism directly as a sort of declarative 
description; we rather borrow some principles and terminology to make the 
foundations of our approach more conventional.

We provide \lstinline{camlp5} syntax extension which generates type-specific
traversal function, abstract transformer and some additional decorations 
from a type declaration. For example

\begin{lstlisting}
    ~\graybox{@}~type lam = 
      Var of string 
    | App of lam * lam
    | Lam of string * lam
\end{lstlisting}

\noindent defines the type itself, its traversal function and abstract transformer 
(from now on we will gray out the constructs, specific to our framework, including 
extended syntax). With this definition we may use predefined type-indexed traversal function 
\lstinline{transform(lam)} and abstract transformer class 
\lstinline{@lam} to implement various concrete transformations.

Suppose we have a polymorphic ADT

\begin{lstlisting}[mathescape]
    type [$\alpha_i$] t = [$C_j$ of [$t^j_{k_j}$]]
\end{lstlisting}

\noindent where we use square brackets to denote a vector of something; here $\alpha_i$ 
denotes $i$-th type parameter, $C_j$ --- $j$-th constructor of the type \lstinline{t}, 
[$t^i_{k_j}$] --- vector of \mbox{$j$-th} constructor argument types. The transformation we are 
looking for has the following type:

\begin{lstlisting}[mathescape]
    $\iota$ $\to$ [$\alpha_i$] t $\to$ $\sigma$
\end{lstlisting}

Here $\iota$ --- the type variable which designates the type of \emph{inherited attribute} 
--- some auxiliary argument which might be helpful to perform the transformation, 
$\sigma$ --- the type variable which designates the type of transformation's result. 
We call this result \emph{synthesized attribute}. Since \lstinline{t} is polymorphic, to 
perform the transformation we might need some transformations of its type arguments. 
Thus our type evolves into

\begin{lstlisting}[mathescape]
    $[\iota \to \alpha_i \to \uparrow\alpha_i]$ $\to$ $\iota$ $\to$ [$\alpha_i$] t $\to$ $\sigma$
\end{lstlisting}

We denote $\uparrow\alpha_i$ the type of synthesized attribute for the transformation of 
$\alpha_i$. Note that we treat type parameter transformations as attribute transformations 
as well; note also that these transformations can provide \emph{different} types 
of synthesized attributes, but operate on the same type of inherited attribute. 
This may look somewhat restrictive; however, we always can ``lift'' different types of 
inherited attributes into their sum type.

The reasoning given above lets us to provide the type signature of 
\emph{abstract transformer}. Abstract transformer for a given type is a  
virtual\footnote{In the context of Objective Caml ``virtual'' means ``purely abstract''.} 
class; all concrete transformation objects for the type are implemented in our 
framework as instances of its subclasses. Abstract transformer is polymorphic over 
the types of inherited and synthesized attributes, type parameters of the transforming 
type (if any) and the types of synthesized attributes for those type parameters. 
Abstract transformers are an 
important part of our solution. Since we are aimed at code reuse based on inheritance 
we can not use implicit transformation objects (which can not be inherited in 
Objective Caml)~\cite{OCaml}; on the other hand polymorphic class definitions
as a rule require precise annotations for the types of their methods. These
type annotations in our case can be quite verbose and tedious to specify. Providing
the single supertype for all concrete transformations allows us not only to
specify the type of the traversal function but also to automatically
instantiate all types of concrete transformation object methods.

The header of the abstract transformer for the type \lstinline{t} looks 
like\footnote{In Objective Caml classes can be polymorphic; type parameter list 
enclosed in a square brackets should precede the name of the class in its 
declaration.}

\begin{lstlisting}[mathescape]
    virtual class [$[\alpha_i$, $\uparrow\alpha_i]$, $\iota$, $\sigma$] ~\graybox{@t}~
\end{lstlisting}

Here the nested pair of square brackets again indicates the vector of pairs of type
variables. Note that we need to know only the vector of type parameters to decide on
the signature of the abstract transformer for that type. For example, in the concrete 
syntax, if we have a type \lstinline{('a, 'b, 'c) t}, then the abstract transformer 
for \lstinline{t} has the following signature:

\begin{lstlisting}
    virtual class ['a, 'ta, 'b, 'tb, 'c, 'tc, 'inh, 'syn] ~\graybox{@t}~
\end{lstlisting}

Here \lstinline{'ta}, \lstinline{'tb}, \lstinline{'tc} designate the types of 
synthesized attributes for the transformations of corresponding type parameters, 
\lstinline{'inh}, \lstinline{'syn} --- types of inherited and synthesized attributes 
for the transformation for \lstinline{t}. By \lstinline{@t} we denote some synthetic 
name for the abstract transformer's class since in Objective Caml classes and types share 
the same namespace.

The next component of our solution is a per-type traversal function which performs pattern
matching and passes control to the transformation object. Taking into account all 
previous type considerations its type should look like

$$
\underbrace{[\iota\to\alpha_i\to\uparrow\alpha_i]}_{\mbox{A}}\to\underbrace{\mbox{\lstinline{[}}[\alpha_i,\uparrow\alpha_i],\iota,\sigma\mbox{\lstinline{] #@t}}}_{\mbox{B}} \to\iota\to [\alpha_i]\;\mbox{\lstinline{t}}\to\sigma
$$

Here $A$ outlines the vector of transformations for the type parameters, $B$ --- the type of 
concrete transformation object, which should be an arbitrary properly instantiated
subtype of abstract transformer \lstinline{@t}\footnote{In Objective Caml, \lstinline{\#t} 
denotes the arbitrary subtype for a class type \lstinline{t}.}. The rest of components 
are inherited attribute type, type of data structure to transform and the type 
of synthesized attribute.

The last thing we need to define is the set of parameters which are passed to the methods 
of transformation object during pattern matching. Since we are dealing with attribute 
transformations we must provide for each method an inherited attribute, which comes as
a parameter to the traversal function. Besides that, each method may need synthesized 
attributes for some (sub)values of the matched value. In our framework synthesized 
attributes can be calculated only by applying some transformation functions. The only
functions available are those for type parameters (supplied as arguments) or the 
transformation for the type of interest. 

The exact implementation of the pattern matching within the traversal function is as 
follows: for the constructor \lstinline[mathescape]{$C$ of $t_i$} we add the 
following case to the matching construct:

\begin{lstlisting}[mathescape]
    | $C$ $[p_i]$ as $s$ ->$\;$ t#c_$C$ i $\bar{s}$ $[\bar{p_i}]$
\end{lstlisting}

Here \lstinline{t} --- transformation object, \lstinline{i} --- inherited attribute 
(both are passed to the traversal function as parameters). $\bar{s}$ and $\bar{p_i}$ 
are \emph{augmented} versions of $s$ (the original node of the transforming data structure) 
and $p_i$ (all proper subvalues of $s$). Namely, we augment these values with functions 
which deliver synthesized attributes when applied to inherited ones. The rules for the 
augmentation are as follows:

\begin{enumerate}
\item if the type of the augmenting value corresponds to a certain type parameter, 
we augment it with the corresponding transformation function for that type parameter;
\item if the type of the augmenting value is \lstinline[mathescape]{$[\alpha_i]$ t}, 
where $\alpha_i$ --- type parameters, \lstinline{t} --- the type we are implementing 
the current traversal function for, then we augment it with the partial application of 
the same traversal function to the transformation functions for corresponding type 
parameters and the same concrete transformation object;
\item in all other cases we do not augment the value.
\end{enumerate}

In cases when we perform the augmentation we also augment the value with the set of all transformation functions for
all type parameters.

The augmented value is represented as a structure type \lstinline{a} with the 
following fields:

\begin{itemize}
\item \lstinline{x} --- the value which was augmented;
\item \lstinline{f} --- the transformation function for the values of the type 
of \lstinline{x};
\item \lstinline{fx} --- the partial application of \lstinline{f} to \lstinline{x};
\item \lstinline{tp} --- the set of transformation functions for all type parameters 
(encoded as object with corresponding methods).
\end{itemize}

We demonstrate these constructs by the following example. Let we have the following type 
definition:

\begin{lstlisting}
    type ('a, 'b) t =
        A of 'a
      | B of 'b
      | T of ('a, 'b) t
\end{lstlisting}

The traversal function for this type is

\begin{lstlisting}
    let rec t_gcata fa fb trans inh subj =
      let rec self = t_gcata fa fb trans
      and tpo = object method a = fa method b = fb end in
      match subj with
        A p1 -> trans#c_A inh (make self subj tpo) (make fa p1 tpo)
      | B p1 -> trans#c_B inh (make self subj tpo) (make fb p1 tpo)
      | T p1 -> trans#c_T inh (make self subj tpo) (make self p1 tpo)
\end{lstlisting}

Since we have two type parameters, the traversal function takes two transformation functions
--- \lstinline{fa} and \lstinline{fb} ---  as its parameters. Then, we need one augmenting 
function for the type itself (\lstinline{('a, 'b) t}). This function is called 
\lstinline{self}. Finally, we need a collection of transformation functions for the type 
parameters encoded by an object with corresponding method names --- hence \lstinline{tpo}. 
The augmenting primitive here is called \lstinline{make}.

Now, when we are implementing the concrete transformation class, we may think in 
terms of inherited and synthesized attributes and attribute transformations. For example, 
writing the method for the constructor \lstinline{A}, say 

\begin{lstlisting}
    method c_A inh s x = ...
\end{lstlisting}

\noindent we know the following:

\begin{itemize}
\item \lstinline{inh} is the inherited attribute;
\item \lstinline{s.f} and \lstinline{x.f} equals to the same transformation 
function we are dealing with now;
\item \lstinline{s.tp#a} is the transformation function for the type 
parameter \lstinline{'a};
\item \lstinline{s.tp#b} is the transformation function for the type 
parameter \lstinline{'b};
\item \lstinline{s.fx} is a function which calculates the synthesized attribute 
for \lstinline{s} with respect to some inherited attribute (for example, but not 
necessary, \lstinline{inh});
\item \lstinline{x.fx} is a function which calculates the synthesized attribute 
for \lstinline{x} with respect to some inherited attribute (for example, but not 
necessary, \lstinline{inh}).
\end{itemize}

Note that due to a late binding for objects the concrete implementations
of augmenting functions can be redefined in subclasses. This property
is important for code reuse.

Despite being rather simple in design the approach in question turned out to 
be tricky in implementation via a syntax extension due to a limited
amount of information available about externally declared types. Another 
problem arises for mutually-recursive type declarations ---
the na\"ive implementation using mutually-recursive classes does not
provide extensible solution since each abstract transformer explicitly 
references abstract transformers for co-recursive types. To provide extensible
solution we had to abstract these transformers by a certain parameterization
which in implementation resulted in dealing with parameterized classes,
mutability, class-level let-bindings and initializers. We omit exact 
description since it is too technical and specific. Finally, polymorphic 
classes in Objective Caml are \emph{regular}, which means that only instances
with the same type parameter bindings can be created within their scopes. 
While this limitation in principle can be worked around using extra 
parameterization via explicitly-polymorphic functions we did not implement
this option yet.

Nevertheless apart from the ``regularity'' limitation (w.r.t. to mutual
recursion) our syntax extension provides complete support for ADTs, polymorphic
variant types~\cite{PolyVar}, structures and tuples, including mutually
recursive type declarations.

The diversity of generic type-driven transformations is widely acknowledged. 
Even for the string conversion functions like \lstinline{show} there are many
various options --- for example, conversion to HTML or XML formats, printer combinators, 
incremental append into string buffer etc. We admit that all these 
cases are rather simple and regular; however, the simpler transformation the 
more distressful it would be to implement it manually for each type of interest. 

Our syntax extension can be customized by the end-user via rather simple plugin 
interface. For example, in the following fragment

\begin{lstlisting}
    ~\graybox{@}~type lam = 
      Var   of string 
    | App   of lam * lam
    | Lam   of string * lam ~\graybox{\underline{with} show}~
\end{lstlisting}

\noindent \lstinline{show} designate plugin name; there are no hardcoded 
transformations in our system at all; ``\lstinline{with}'' construct 
plays role of plugin invocation primitive. 

Each plugin is dynamically loaded during the syntax extension phase and generates 
concrete transformation on a per-type (actually, per-constructor) basis. Since any 
transformation in our framework is represented by a certain class, each plugin  
actually generates one class per type. To address plugin-defined transformation
\lstinline{p} for the type \lstinline{t} extended construct 
\lstinline[mathescape]{$\graybox{@p[t]}$} can be used. 
Most work is performed by the core system; the plugin 
itself provides rather simple parameterization plus a concrete function to generate 
the body for each method. For example, the implementations of \lstinline{show} 
contains less than 50 lines, about 1/3 of which are just interface 
ceremonial code. 

\section{Examples}

In this section we present some use cases which we believe demonstrate
the potential of our framework in terms of code reuse. 

\subsection{Extensible \lstinline{fold}}

Folds, or catamorphisms~\cite{Fold,Bananas}, represent a wide class
of useful transformations. We introduce here an extensible generic
fold which can be used to derive these transformations.

For our main example we may implement the following transformation
class:

\begin{lstlisting}
    class ['a] fold_lam = object inherit ['a, 'a] ~\graybox{@lam}~
      method c_Var s _ _   = s
      method c_App s _ x y = ~\graybox{x.fx (y.fx s)}~
      method c_Lam s _ x y = ~\graybox{y.fx s}~
    end
\end{lstlisting}

This implementation looks completely vacuous at the first glance: \lstinline{fold_lam} 
simply threads the inherited attribute through all the nodes and finally returns 
it untouched. However, this behavior is just what we need as a basis for various 
transformations which can be obtained by a proper modification. For example,

\begin{lstlisting}
    class vars = object inherit [S.t] fold_lam
      method c_Var s _ x = S.add x s
    end
\end{lstlisting}

\noindent gives us the set of all variables occurred in a lambda-term (here 
\lstinline{S.t} stands for the type of string sets). The set of all 
free variables can be calculated by inheriting from the class \lstinline{vars}:

\begin{lstlisting}
    class free_vars = object inherit vars   
      method c_Lam s _ x l = S.union s (S.remove x (~\graybox{l.fx S.empty}~))
    end 
\end{lstlisting}

As we can see, in the implementation of \lstinline{vars} we reused two cases from
\lstinline{fold_lam}, and in the implementation of \lstinline{free_vars} we
(again) reused two cases from \lstinline{vars}. Without late binding we
would have to provide two different functions with complete case analysis to
fold with.

The similar considerations are applicable to yet another important
transformation --- ``\lstinline{map}''. Indeed, having a ``default''
implementation in the form of copying we may then redefine
its behavior for the ``interesting'' cases providing various
useful concrete transformations. Predefined plugins for \lstinline{fold} and 
\lstinline{map} are included in our framework among \lstinline{show}.

\subsection{``Expression Problem''}

``Expression problem''~\cite{ExpressionProblem} is a widely recognized
reference task in the area of component-based software development. The
task is to implement an expression evaluator which can be incrementally 
extended with new cases without modifications of existing code. 

Expression problem can easily be solved in Objective Caml using polymorphic
variants~\cite{PolyVarReuse}. Polymorphic variant types~\cite{PolyVar} 
are an extended version of regular ADTs which was introduced in Objective 
Caml starting from the version 3. In short, unlike regular ADTs, 
for which a certain constructor can belong to exactly one type (in its scope), 
different polymorphic variant types can share the same constructors. 
This, in particular, creates a possibility to operate with types in a 
structural, not nominal, way. For example, polymorphic variants can 
be subtyped, inherited and defined implicitly.

Here we demonstrate that our framework can provide more declarative 
implementation for the same solution as in~\cite{PolyVarReuse} (see 
Fig.~\ref{ExpressionProblemSolution}).

\begin{figure}[t]
\begin{center}
\begin{minipage}{0.8\linewidth}
\begin{lstlisting}[basicstyle=\scriptsize]
    ~\graybox{\bfseries @}~type var = [`Var of string] 
    class ['v] var_eval = object inherit [string -> 'v, 'v] ~\graybox{@var}~
      method c_Var s _ x = s x
    end

    ~\graybox{\bfseries @}~type 'a arith = [`Add of 'a * 'a | `Mul of 'a * 'a] 
    class ['a, 'b] arith_eval = object inherit ['a, int, 'b, int] ~\graybox{@arith}~
      method c_Add s _ l r = ~\graybox{l.fx s}~ + ~\graybox{r.fx s}~
      method c_Mul s _ l r = ~\graybox{l.fx s}~ * ~\graybox{r.fx s}~
    end

    ~\graybox{\bfseries @}~type 'a expr = [ var | 'a arith ] 
    class ['a] expr_eval = object
      inherit ['a, int, string->int, int] ~\graybox{@expr}~
      inherit [int] var_eval
      inherit ['a, string -> int] arith_eval
    end

    let rec eval s e = transform(expr) eval (new expr_eval) s e
\end{lstlisting}
\end{minipage}
\end{center}
\caption{Solution for an instance of the Expression Problem}
\label{ExpressionProblemSolution}
\end{figure}

First, we declared two partial expression types (\lstinline{var} and
\lstinline{arith}) with their evaluators (\lstinline{var_eval} and
\lstinline{arith_eval}); type \lstinline{arith} is made polymorphic in advance to be
more open --- this is the feature of the original solution. Note that these 
evaluators are polymorphic: in the case of type \lstinline{var} we do 
not know the type of evaluation result (synthesized attribute, \lstinline{'v}); 
in the case of type \lstinline{arith} we do not know the type of state 
(inherited attribute, \lstinline{'b}). This property ensures us that 
we did not introduce any artificial restrictions for these evaluators. 
We then combine both partial types (via regular inheritance for polymorphic 
variants) and their evaluators (via regular inheritance for classes). 
Now we can unambiguously determine the types of state (\lstinline{string->int}) 
and evaluation result (\lstinline{int}). The type \lstinline{expr} is again
open, so we have to ``tie the knot'' in the top-level evaluator 
(\lstinline{eval}).

In our implementation all ``glue code'', needed in~\cite{PolyVarReuse} to combine 
the evaluators for the partial types, is generated by the framework and completely 
invisible on the user level.

\subsection{Lambda Calculus Reductions}

For the final example we consider the implementation of different reduction
strategies for lambda calculus. As a reference we choose Peter Sestoft's 
paper~\cite{Sestoft}, which provides a nice categorization of reduction order
steps. Seven reduction strategies are described using this categorization 
in terms of big-step operational semantics; for three of them a reference 
ML-implementation is provided. Here we present an implementation which 
literally follows Sestoft's reasonings.

\begin{figure}[t]
\begin{center}
\begin{minipage}{0.8\linewidth}
\begin{lstlisting}[basicstyle=\scriptsize]
    type context = string -> string
    type mtype   = context -> ~\graybox{(context, lam, lam, < >) a}~ -> lam

    class virtual reducer = object(this) inherit [context, lam] ~\graybox{@lam}~
      method virtual arg       : mtype
      method virtual subst_arg : mtype
      method         head      : mtype = fun c x -> ~\graybox{x.fx c}~
      method c_Var _ x _   = ~\graybox{x.x}~
      method c_App c s l m = match this#head c l with
      | Lam (x, l') -> ~\graybox{s.f c (subst c x (this\#subst\_arg c m) l')}~
      | l'          -> let l'' = ~\graybox{s.f c l'}~ in 
                      App (l'', this#arg c m)
    end
\end{lstlisting}
\end{minipage}
\end{center}
\caption{Implementation of lambda reductions: base class}
\label{LRBase}
\end{figure}

The type of lambda expressions we have been considering so far was actually 
borrowed from the referenced paper; for this type we can declare the following
virtual class which abstracts any admissible reduction order according
to Sestoft's categorization (see Fig.~\ref{LRBase}). Apart from inheriting from 
the abstract transformer \lstinline{@lam} we introduce the following
supplementary methods:

\begin{itemize}
\item Method \lstinline{head} reduces a lambda expression
in a head position; we provide here a default implementation
which uses exactly the same reduction order as being defined. 
However this is not always the case --- some reduction
strategies (called ``\emph{hybrid}'' in~\cite{Sestoft}) use different 
orders for this purpose.
\item Method \lstinline{arg} reduces a lambda expression in
an argument position when expression in corresponding
head position was not reduced to abstraction.
\item Method \lstinline{subst_arg} reduces a lambda expression in
an argument position when expression in corresponding head position was 
reduced to abstraction.
\end{itemize}

Type \lstinline{context} represents name-generating function
needed to perform alpha-conversions and plays role of the
inherited attribute. Type \lstinline{mtype} used as an abbreviation
for types of supplementary methods, which transform inherited attribute 
(\lstinline{context}) and augmented lambda expression (see Section~\ref{elaborated})
into lambda expression. Finally we provide two generic 
implementations for reducing variables and applications. Variables
never reduced; in application the expression in a head
position is reduced first with method \lstinline{head}, then the result
is inspected: if it is an abstraction then its argument is reduced
with \lstinline{subst_arg} and then a substitution is performed, otherwise 
its argument is reduced with \lstinline{arg}. We did not include in the 
listing the implementation of substitution function \lstinline{subst}.

Given this base class we can define various traits of reduction orders (see
Fig.~\ref{LRTraits}); to discriminate cases we explicitly follow~\cite{Sestoft}. 

\begin{figure}
\begin{tabular}{p{6cm}p{6cm}}
\begin{lstlisting}[basicstyle=\scriptsize]
class virtual reduce_under_abstractions = 
  object inherit reducer
    method c_Lam c _ x l = Lam (x, ~\graybox{l.fx c}~)
  end

class virtual reduce_arguments = 
  object inherit reducer 
    method arg c x = ~\graybox{x.fx c}~
  end

class virtual non_strict = 
  object inherit reducer
    method subst_arg _ m = ~\graybox{m.x}~
  end
\end{lstlisting} &
\begin{lstlisting}[basicstyle=\scriptsize]
class virtual dont_reduce_under_abstractions = 
  object inherit reducer 
    method c_Lam _ s _ _ = ~\graybox{s.x}~
  end

class virtual dont_reduce_arguments = 
  object inherit reducer
    method arg _ x = ~\graybox{x.x}~
  end

class virtual strict = 
  object inherit reducer
    method subst_arg c m = ~\graybox{m.fx c}~
  end
\end{lstlisting}
\end{tabular}
\caption{Implementation of lambda reductions: reduction order traits}
\label{LRTraits}
\end{figure}

Having all these traits defined we finally can implement all
reduction orders by simply combining relevant traits via inheritance. Some
interesting cases are presented on Fig.~\ref{LR}. Note that this example 
demonstrates that object representation of transformations provides
more then just componentization --- in cases of applicative
order or hybrid applicative order we
did not combine the reduction transformations ``from scratch'' using the
basic traits; we rather redefined some traits in already completely 
implemented transformations\footnote{In the latter case the order of inheritance 
clauses is important since each inheritance (re)binds some methods.}.

In order to trace individual reductions during normalization a second 
implementation is sketched in the original paper. In that implementation 
lambda-term under reduction is represented as a redex and its 
context --- lambda-term with a hole. Contexts represented by a 
functions with type \lstinline{lam -> lam}. A small set of combinators 
is provided to manipulate context which is passed as auxiliary 
argument and modified as reduction point advances into the original term. 
Since this modification is global it affects the implementation of each 
reduction strategy. 

In our case, however, it is sufficient to modify only the implementations 
of the base class and one reduction order trait. The definitions
of all other traits and individual reduction strategies can be 
left completely intact. The complete self-contained implementation for 
both cases is included into Appendix.

\begin{figure}
\begin{tabular}{p{5.8cm}p{7cm}}
\begin{lstlisting}[basicstyle=\scriptsize]
class call_by_name = object 
  inherit dont_reduce_under_abstractions 
  inherit dont_reduce_arguments
  inherit non_strict
end 
let bn = transform(lam) (new call_by_name)

class applicative = object
  inherit call_by_value
  inherit reduce_under_abstractions
end
let ao = transform(lam) (new applicative)
\end{lstlisting}&
\begin{lstlisting}[basicstyle=\scriptsize]
class call_by_value = object
  inherit dont_reduce_under_abstractions
  inherit reduce_arguments
  inherit strict
end
let bv = transform(lam) (new call_by_value)

class hybrid_applicative = object
  inherit applicative
  method head c x = bv c x.x
end
let ha = transform(lam) (new hybrid_applicative)
\end{lstlisting}
\end{tabular}
\caption{Implementation of lambda reductions: some of reduction orders}
\label{LR}
\end{figure}

\section{Related Works}

Our approach was specifically designed for Objective Caml; it is
interesting to discuss other languages it can be implemented for. 
Such languages must at least combine first-class functions, objects 
with inheritance and late binding and algebraic datatypes. 
Scala\footnote{http://www.scala-lang.org} and Kotlin\footnote{http://kotlin.jetbrains.org}
can be considered as relevant examples. However in these languages
object-oriented and functional features are not orthogonal --- first-class
functions and ADTs are mimicked and projected into object-oriented layer
which leads to internal object-oriented data representation and makes our
approach superfluous (though possible). As another candidate we can mention
Haskell, in which object-oriented extension can be implemented using 
typeclass-level metaprogramming~\cite{OOHaskell}.

Code reuse problem which we addressed can be dealt with in many various ways.
Datatype-generic programming~\cite{DGP} aimed at reuse of type-driven
transformations; we can mention ``Scrap Your Boilerplate''~\cite{SYB,SYB1,SYB2}
or ``instant generics''~\cite{InstantGenerics} as examples. Some of their
functionality (e.g., serialization, generic maps and queries) definitely 
can be reproduced using our framework in terms of plugins. 
In~\cite{Bananas} a general constructive approach is provided for various classes of 
type-driven transformations. From that perspective all transformations we are able to 
implement using our framework can be characterized as \emph{catamorphisms}. 
Recursive types considered as fixed points of (poly)functors; any catamorphism 
can be represented via combination of type-specific traversal function (``fold'')
and transformation-specific function (``algebra''). There is a strong analogy
between this approach and ours --- ``fold'' corresponds to our type-indexed
``transform'', while ``algebra'' --- to object-encoded transformer. The difference
is that our ``transform'' is non-recursive, concrete transformation
functions are captured and passed to object-encoded transformer as 
parameters. So, in our framework the traversal order of data structure can 
be made specific not to its type, but to a concrete transformation; so 
generally we can implement a superset of catamorphism class described 
in~\cite{Bananas}.

Another relevant approach which in fact motivated our work is suggested 
in~\cite{AGSwierstra}. Attribute grammar-based domain-specific language is proposed 
for describing catamorphisms in a declarative form. This domain-specific language 
is implemented as a preprocessor for Haskell. A specifications for different
attributes and their evaluation rules can incrementally be introduced, modified 
and combined yielding transformation which performs all these evaluation 
provided these descriptions are consistent with each other. The consistency 
property is statically checked using advanced type-system encoding including type 
arithmetics.

The direct reuse of the aforementioned approach is impossible due to the differences 
in type systems between Haskell and Objective Caml (Haskell implementation utilizes 
heterogeneous collections and, hence, requires type arithmetic). Moreover, the explicit 
encoding of attributes and their evaluation rules can compromise the idea of code 
reuse since we cannot have \emph{different evaluation rules} for the \emph{same attribute} 
(as we actually had two \emph{different} \lstinline{show} functions for the \emph{same} 
type in Section~\ref{overview}). 

Another approach to code reuse is
concentrated on composing software from separately developed reusable
components. We already mentioned ``Expression problem''\cite{ExpressionProblem} 
as a reference task in this area. ``Expression problem'' can be solved 
in a number of languages including Haskell~\cite{ALaCarte}, 
Java~\cite{ObjectAlgebras} and Objective Caml~\cite{PolyVarReuse}. As we demonstrated,
our framework is compatible with the solution for Objective Caml; moreover it
allows even more code reuse since with our framework it is possible to
modify the behavior of completely assembled transformation (which is impossible,
for example, in~\cite{ALaCarte} due to uniqueness of class membership instantiation
for a given type).

From the technical point of view our approach resembles 
\lstinline{MapGenerator}\footnote{http://brion.inria.fr/gallium/index.php/Camlp4MapGenerator}
and \lstinline{FoldGenerator}\footnote{http://brion.inria.fr/gallium/index.php/Camlp4FoldGenerator} which are shipped with \lstinline{camlp4}. These tools provide syntax extensions 
which generate ``map'' and ``fold'' transformations in a form of a single class per a 
cluster of mutually-recursive type definitions. Each method of this class represents the 
transformation for each type of the cluster. This representation indeed allows to modify 
the behavior of transformation by inheritance. There are, however, some differences, which
make that approach less general then ours:

\begin{enumerate}
\item Method-per-type representation does not eliminate the need for pattern-matching. 
Extending transformation method one still needs to manually match against 
``interesting'' constructors which we consider a boilerplate.

\item Similarly, the transformation for a union of polymorphic variant types can not 
be constructed by inheritance from the transformations for its counterparts --- 
some ``glue'' code is needed.

\item There are many other interesting transformations which are left overboard 
(e.g. ``show'', ``compare'' etc.). While some of them technically can be seen as 
specializations of fold or map in fact no code reuse can be achieved by utilizing 
\lstinline{Fold/MapGenerator} since the body of each generated method has to 
be completely reimplemented. In our framework these transformation can be
implemented in form of plugins with short and simple implementation.
\end{enumerate}

Another ``close relative'' of our framework is ``deriving'' syntax extension and 
library\footnote{https://code.google.com/p/deriving}. Like our plugin system, 
``deriving'' allows to generate specific functionality from type definitions, 
and the assortment of traits can potentially be extended by the end-user~\cite{Yallop}. 
However this framework does not utilize object-encoding which makes the generated 
traits less flexible.

\section*{Conclusions}

We presented a generic programming framework for Objective Caml which is based on the 
notion of object-encoded transformations. Proposed approach facilitates even more code 
reuse in comparison with conventional tools such as classes/objects and polymorphic 
variants since it allows to combine best practices from both functional and 
object-oriented programming. The implementation consists of a syntax extension which 
operates on slightly augmented descriptions of Objective Caml types and very small 
runtime library. Additionally the framework itself can be parameterized by mean of 
plugins which provide functionality to generate custom type-driven transformations 
conforming the generic framework interface. We believe that a wide range of 
transformations for regular ADTs and polymorphic variant types can be expressed 
using approach in question. Finally, our approach gives a good example of object-oriented 
programming in terms of Objective Caml.

Recent development of Objective Caml introduced some new features such as GADT, open
types and extensible functions. An improvement of our implementation to support 
these features can be considered as future work as well as elimination of
``regularity'' restriction. As another important problem we can mention performance
evaluation of generic vs hand-written transformations and reducing
genericity-imposed overhead.

\clearpage

\section*{Appendix: Lambda Calculus Reduction Orders}

\begin{lstlisting}[commentstyle=\scriptsize\ttfamily\itshape,basicstyle=\scriptsize]
(* Opening runtime module to avoid explicit qualification *)
open GT

(* Set/map of strings *)
module S = Set.Make (String) 
module M = Map.Make (String) 

(* Name generator for alpha-conversion *)
let generator s = 
  let s = ref s in
  let rec gen x =
    let n = x ^ "'" in
    if S.mem n !s 
    then gen n
    else (s := S.add n !s; n)
  in gen

(* Helper class for alpha-conversion; parameterized
   by name-generating function ``g'' and a set of ``prohibited''
   free variables ``fvs''
*)
class substitutor gen fvs =
  object (this)
    val s = (M.empty : string M.t)
    method subst  x = try M.find x s with Not_found -> x
    method rename x = if S.mem x fvs 
                      then let x' = gen x in x', {< s = M.add x x' s >} 
                      else x, {< >}
  end

(* Type of lambda expression; enables ``show'' and ``foldl'' transformations *)
@type lam = 
| Var of string 
| App of lam * lam
| Lam of string * lam with show, foldl

(* ``show'' function *)
let show = transform(lam) (new @show[lam]) ()

(* Transformation class to collect variables; reuses ``foldl'' *)
class var = object inherit [S.t] @foldl[lam]
  method c_Var s _ x = S.add x s
end

(* Function to collect variable names *)
let vars = transform(lam) (new var) S.empty

(* Context --- function to generate ``fresh'' names *)
let context l = generator (vars l)

(* Transformation class to collect free variables; reuses ``var'' *)
let fv = transform(lam) 
  object inherit var   
    method c_Lam s _ x l = S.union s (S.remove x (l.fx S.empty))
  end S.empty 

(* Substitution function (generic as well) *)
let subst g x m = transform(lam) 
  object inherit [substitutor, lam] @lam
    method c_Var s _ y = 
      if y = x then m else Var (s#subst y)
    method c_Lam s z y l = 
      if y = x then z.x
               else let y', s' = s#rename y in Lam (y', l.fx s')
    method c_App s _ l m = App (l.fx s, m.fx s)
  end (new substitutor g (fv m))

(* Module type to abstract base class for implementing reduction orders *)
module type Reducer =
  sig

    (* Abstract type for inherited attribute *)
    type context
    (* Shortcut for augmented type of lambda-expression *)
    type aug    = (context, lam, lam, < >) a
    (* Shortcut for type of supplementary methods *)
    type mtype  = context -> aug -> lam

    (* Abstract function to provide ``default'' inherited attribute *)
    val default : lam -> context 

    (* Template base class for reduction trnsformation *)
    class virtual reducer :
      object inherit [context, lam] @lam
        method virtual arg       : mtype
        method virtual subst_arg : mtype
        method         head      : mtype 
        method         c_Var     : context -> aug -> string -> lam
        method         c_App     : context -> aug -> aug -> aug -> lam
      end

    (* Template class for the only trait which is sensitive to 
       the ``context'' type
    *)
    class virtual reduce_under_abstractions :
      object inherit reducer
      method c_Lam : context -> aug -> string -> aug -> lam
    end

  end

(* Functor to implement concrete reduction orders *)
module Reductions (R : Reducer) = 
  struct
    (* Opening R to avoid qualifications *)
    open R

    (* Top-level reduction function: applies reduction
       order ``r'' to lambda-term ``l''
    *)
    let reduce r l = r (default l) l

    (* Basic reduction order traits *)
    class virtual dont_reduce_under_abstractions = 
      object inherit reducer 
      method c_Lam _ s _ _ = s.x
    end

    class virtual reduce_arguments =
      object inherit reducer 
        method arg c x = x.fx c
      end

    class virtual dont_reduce_arguments =
      object inherit reducer
        method arg _ x = x.x
      end

    class virtual non_strict =
      object inherit reducer
        method subst_arg _ m = m.x
      end

    class virtual strict =
      object inherit reducer
        method subst_arg c m = m.fx c
      end

    (* Reduction orders *)
    class call_by_name = object 
      inherit dont_reduce_under_abstractions 
      inherit dont_reduce_arguments
      inherit non_strict
    end 

    let bn = transform(lam) (new call_by_name)

    class normal = object
      inherit reduce_under_abstractions
      inherit reduce_arguments
      inherit non_strict
      method  head c x = bn c x.x
    end 
    let nor = transform(lam) (new normal)

    class call_by_value = object
      inherit dont_reduce_under_abstractions
      inherit reduce_arguments
      inherit strict
    end
    let bv = transform(lam) (new call_by_value)

    class applicative = object
      inherit call_by_value
      inherit reduce_under_abstractions
    end
    let ao = transform(lam) (new applicative)

    class hybrid_applicative = object
      inherit applicative
      method head c x = bv c x.x
    end
    let ha = transform(lam) (new hybrid_applicative)

    class head_spine = object
      inherit call_by_name
      inherit reduce_under_abstractions
    end 

    let he = transform(lam) (new head_spine)

    class hybrid_normal = object
      inherit normal 
      method  head c x = he c x.x
    end
    let hn = transform(lam) (new hybrid_normal)

    (* Top-level definitions *)
    let sample r l =
      Printf.printf "%s ----> %s\n" (show l) (show (r l)) 

    let main () =
      let run n r =
        Printf.printf "\n========== %s ================\n\n" n;
        List.iter (sample r) [
          Lam ("x", App (Lam ("y", Var "y"), Var "z"));
          App (Lam ("x", App (Lam ("y", Var "y"), Var "z")), Var "y");
          App (Var "x", App (Lam ("x", Var "x"), Var "y"));
          App (Lam ("x", App (Var "x", Var "y")), App (Lam ("x", Var "x"), Var "y"));
          App (Lam ("x", App (Var "y", Var "x")), App (Lam ("x", Var "x"), Var "y"));
        ] 
      in
      run "Call-by-name"        (reduce bn);
      run "Normal Order"        (reduce nor);
      run "Call-by-value"       (reduce bv);
      run "Applicative"         (reduce ao);
      run "Hybrid Applicative"  (reduce ha);
      run "Head Spine"          (reduce he);
      run "Hybrid Normal Order" (reduce hn)

  end

(* Top-level definition *)
let _ =
  (* Simple case --- reduction with no context tracing *)
  let module Simple = Reductions (
    struct
      (* Inherited attribute: name-generation function *)
      type context = string -> string
      type aug    = (context, lam, lam, < >) a
      type mtype  = context -> aug -> lam

      let default = context

      (* Base reducer for the simple case *)
      class virtual reducer = 
        object(this) inherit [context, lam] @lam
          method virtual arg       : mtype
          method virtual subst_arg : mtype
          method         head      : mtype = fun c x -> x.fx c
          method c_Var _ x _   = x.x
          method c_App c s l m = 
            match this#head c l with
            | Lam (x, l') -> s.f c (subst c x (this#subst_arg c m) l')
            | l'          -> let l'' = s.f c l' in 
                             App (l'', this#arg c m)
         end

      (* Context-type-sensitive trait for the simple case *)
      class virtual reduce_under_abstractions = 
        object inherit reducer
          method c_Lam c _ x l = Lam (x, l.fx c)
        end
    end
  ) 
  in
  (* Advanced case with context reconstruction; the definitions
     of all but one trait and all reduction orders completely reused 
  *)
  let module WithContext = Reductions (
    struct
      (* Inherited attribute: name-generating function and term with a hole *)
      type context = (string -> string) * (lam -> lam)
      type aug     = (context, lam, lam, < >) a
      type mtype   = context -> aug -> lam

      (* Combinators to manipulate terms with holes *)
      let (@@) f g x = f (g x)
      let id   x     = x
      let abst x  e  = Lam (x, e)
      let appl e1 e2 = App (e1, e2)
      let appr e2 e1 = App (e1, e2)
  
      let default l = (context l, id)

      (* Base reducer with context reconstruction *)
      class virtual reducer = 
        object(this) inherit [context, lam] @lam
          method virtual arg       : mtype
          method virtual subst_arg : mtype
          method         head      : mtype = fun c x -> x.fx c
          method c_Var _ x _ = x.x
          method c_App ((g, c) as i) s l m = 
            match this#head (g, c @@ appl l.x) l with
            | Lam (x, l') -> s.f i (subst g x (this#subst_arg (g, c @@ appr l') m) l')
            | l'          -> let l'' = s.f (g, c @@ appl l.x) l' in 
                             App (l'', this#arg (g, c @@ appr l'') m)
         end

      (* Context-type-sensitive trait with context reconstruction *)      
      class virtual reduce_under_abstractions = 
        object inherit reducer
          method c_Lam (g, c) _ x l = Lam (x, l.fx (g, c @@ abst x))
        end

    end
  )
  in
  (* Running both cases *)
  Simple.main ();
  WithContext.main ()
\end{lstlisting}

\begin{thebibliography}{99}
\bibitem{PolyVar}
Jacques Garrigue. Programming with Polymorphic Variants. ICFP Workshop on ML, 1998.

\bibitem{PolyVarReuse}
Jacques Garrigue. Code Reuse Through Polymorphic Variants. FOSE-2000.

\bibitem{OCaml}
Didier R{\'{e}}my. Using, Understanding, and Unraveling the OCaml Language.
Applied Semantics. Advanced Lectures. LNCS 2395, 2002.

\bibitem{Bananas}
Erik Meijer, Maarten Fokkinga, Ross Paterson. Functional Programming with Bananas, Lenses, 
Envelopes and Barbed Wire. 5th ACM Conference on Functional Programming Languages and 
Computer Architecture, 1991.

\bibitem{AGKnuth}
Donald E. Knuth. Semantics of Context-Free Languages.
Mathematical Systems Theory, Vol. 2, No. 2, 1967.

\bibitem{AGSwierstra}
Marcos Viera, S. Doaitse Swierstra, Wouter Swierstra.
Attribute Grammars Fly First-Class: How to do Aspect Oriented Programming in Haskell.
ICFP-2009.

\bibitem{Yallop}
Jeremy Yallop. 
Practical Generic Programming in OCaml. 
ICFP Workshop on ML, 2007.

\bibitem{DGP}
Jeremy Gibbons. Datatype-generic Programming // Spring School on Datatype-Generic 
Programming. LNCS 4719, 2006.

\bibitem{ALaCarte}
Wouter Swierstra. Data Types \`a la Carte // Journal of Functional Programming, Vol. 18, 
No. 4, 2008.

\bibitem{Sestoft}
Peter Sestoft. Demonstrating Lambda Calculus Reductions // The Essence of Computations, LNCS Vol.~2566, 2002.

\bibitem{Fold}
Graham Hutton. A Tutorial on the Universality and Expressiveness of fold // 
Journal of Functional Programming, Vol.~9, No.~4, 1999.

\bibitem{ExpressionProblem}
Philip Wadler, et al. The Expression Problem. Discussion on the Java-Genericity
mailing list, December 1998.

\bibitem{OOHaskell}
Oleg Kiseliov, Ralf L\"ammel. Haskell's Overlooked Object System // arXiv:cs/0509027, 2005.

\bibitem{SYB}
Ralf L\"ammel, Simon Peyton Jones.
Scrap Your Boilerplate: A Practical Design Pattern for Generic Programming.
Workshop on Types in Language Design and Implementation, 2003.

\bibitem{SYB1}
Ralf L\"ammel, Simon Peyton Jones.
Scrap More Boilerplate: Reflection, Zips, and Generalised Casts.
ICFP-2004.

\bibitem{SYB2}
Ralf L\"ammel, Simon Peyton Jones.
Scrap Your Boilerplate with Class: Extensible Generic Functions.
ICFP-2005.

\bibitem{InstantGenerics}
Manuel M. T Chakravarty, Gabriel C. Ditu, Roman Leshchinskiy.
Instant Generics: Fast and Easy, 2009.

\bibitem{ObjectAlgebras}
Bruno C. d. S. Oliveira, William R. Cook.
Extensibility for the Masses: Practical Extensibility with Object Algebras //
ECOOP, 2012.
\end{thebibliography}
\end{document}